\documentclass[aps,prl,twocolumn,superscriptaddress,amsmath,amssymb,longbibliography,footinbib]{revtex4-2}

\usepackage{graphicx}
\usepackage{bm}
\usepackage{braket}
\usepackage{physics}
\usepackage{hyperref}
\usepackage{xcolor}
\newcommand{\bs}{\boldsymbol}
\usepackage{tikz}
\usetikzlibrary{arrows.meta,calc}

\newcommand{\hc}{\mathrm{h.c.}}

\begin{document}

\title{Photon condensation driven by scalar spin chirality
%Superradiant phase transition in triangular molecular rings
}

\author{Dimitrije Stepanenko}
\affiliation{Nanostructured Matter Laboratory, Center for Solid State Physics and New Materials, Institute of Physics Belgrade, University of Belgrade, Pregrevica 118, 11080 Belgrade, Serbia}

\author{Daniel Loss}

\affiliation{Physics Department, King Fahd University of Petroleum and Minerals, 31261, Dhahran, Saudi Arabia}
\affiliation{Quantum Center and RDIA Chair
in Quantum Computing, KFUPM, Dhahran}
\affiliation{Department of Physics, University of Basel, Klingelbergstrasse 82, 4056 Basel, Switzerland}

\author{Mircea Trif}
%\affiliation{Department of Physics, University of Basel, Klingelbergstrasse 82, 4056 Basel, Switzerland}
\affiliation{International Research Centre MagTop, Institute of Physics, Polish Academy of Sciences, Aleja Lotnikow 32/46, PL-02668 Warsaw, Poland}
\date{\today}

\begin{abstract}
Gauge invariance forbids equilibrium photon condensation
for spatially uniform electric-dipole coupling, motivating
alternative routes based on orbital magnetic response.
We show that triangular molecular magnets coupled to
a cavity support photon condensation driven by emergent
scalar spin chirality. Starting from the half-filled
Hubbard model in the Mott regime, we derive an effective
Hamiltonian in which virtual ring exchange couples the
cavity flux directly to scalar spin chirality.
Mean-field theory, collective-fluctuation calculations,
and finite-size exact diagonalization reveal a phase
with spontaneous chirality ordering and a macroscopic
cavity displacement, together with hybrid
chirality-photon excitations.
The condensation threshold can be tuned electrically
through the molecular chirality splitting.
Our results establish emergent orbital degrees of freedom
in frustrated magnets as a route to equilibrium
light-matter ordering beyond conventional dipolar
cavity quantum electrodynamics.
%Equilibrium superradiant phase transitions are forbidden in standard cavity quantum electrodynamics with dipolar light-matter coupling due to gauge invariance. Coupling cavity photons to orbital currents through Peierls phases opens a distinct route toward superradiance in correlated quantum matter. We demonstrate this mechanism for triangular molecular magnets coupled to a cavity, where strong correlations promote a superradiant phase with spontaneous scalar spin chirality. Combining mean-field theory with exact diagonalization of the full electron-photon Hamiltonian, we establish the emergence of a macroscopic cavity field beyond a critical light-matter coupling. In the Mott regime at half filling, we derive an effective spin Hamiltonian in which virtual ring-exchange processes generate the light–chirality coupling, revealing the microscopic origin of the ordered phase. Our work shows how orbital light-matter coupling and geometrical frustration cooperate to realize superradiance in correlated quantum materials, with scalar spin chirality as the underlying electronic order parameter.
\end{abstract}

\maketitle

Equilibrium superradiant or photon-condensed phases
in cavity quantum electrodynamics have long been
constrained by gauge-invariance arguments.
For spatially uniform electric-dipole coupling,
the cancelation between paramagnetic and diamagnetic
terms forbids a macroscopic ground-state photon
occupation—the superradiant no-go theorem
\cite{NoGoPRL75,NoGoPRA79,NoGoPRA81,NatafNC10,AndolinaPRB19}.
This restriction motivates alternative coupling
mechanisms that allow photon condensation without
external driving or dissipation.

\begin{figure}[t]
\centering
\includegraphics[width=\linewidth]{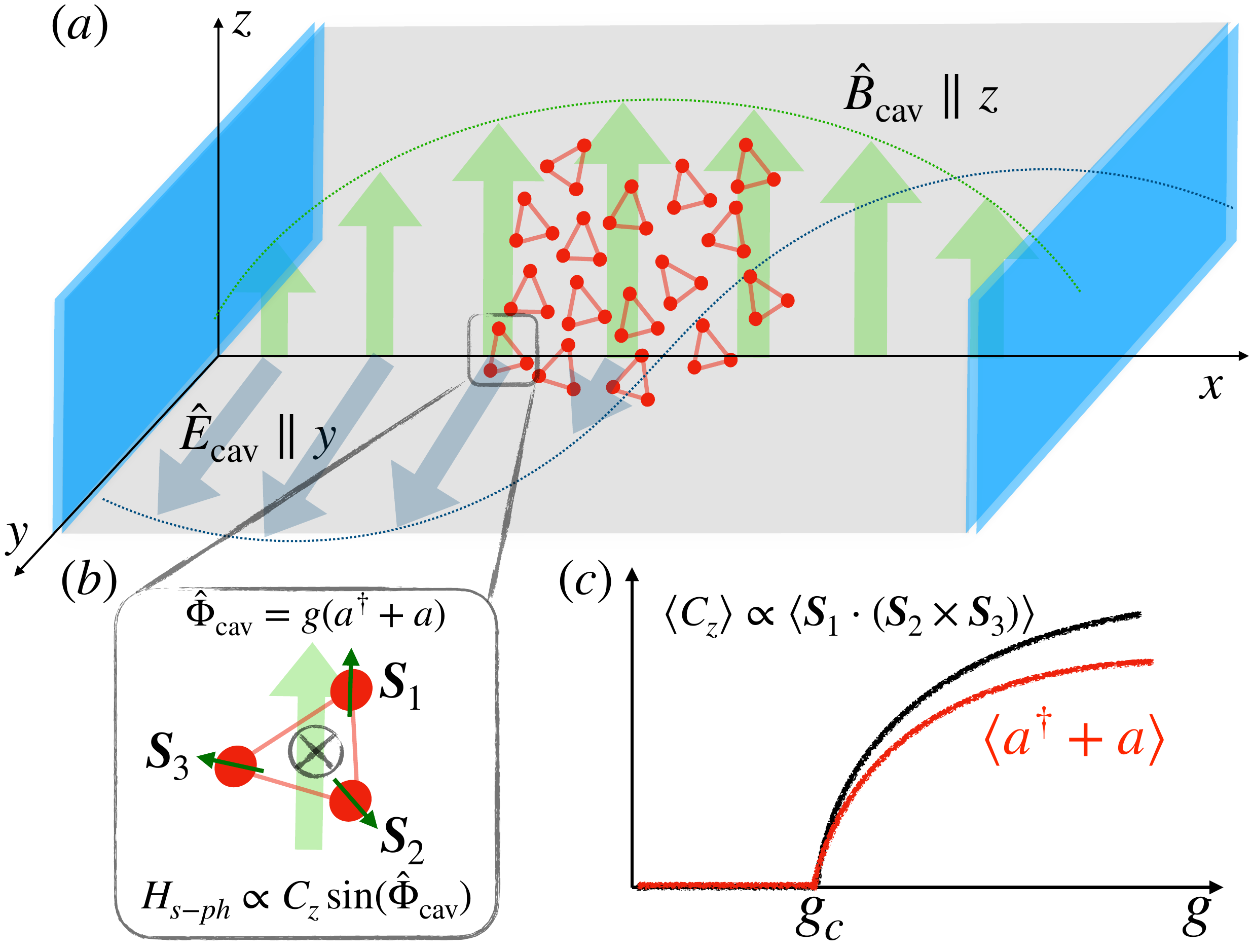}
\caption{Cavity-coupled molecular rings. $(a)$ A standing-wave microwave mode with molecular triangles at the magnetic field $\hat{\boldsymbol B}_{\text{cav}}$ antinode (green) and electric field $\hat{\boldsymbol  E}_{\text{cav}}$ node (gray) to maximize inductive coupling. $(b)$ Individual ring detail: an out-of-plane flux $\hat{\Phi}_{\rm cav}$ couples the cavity to the scalar spin chirality $C_z$ with strength $g \propto|{\boldsymbol B}_{\text{cav}}|A$, with $A$ the area of the molecule (assumed, for simplicity, to lie in the $x-y$ plane). $(c)$ Spontaneous symmetry breaking for $g > g_c$, where a macroscopic photon displacement $\langle a + a^\dagger \rangle$ emerges alongside a finite scalar spin chiral order parameter $\langle C_z \rangle$.
}
\label{fig1}
\end{figure}
Theoretical work has established that spatially varying
cavity fields can evade the uniform-field no-go theorem
and support equilibrium photon condensation through
the orbital magnetic response of electrons
\cite{NatafPRL19,AndolinaPRB20,GuerciPRL20,KozinPRBSC25}.
Unlike a uniform vector potential, a cavity field
carrying magnetic flux modifies the electronic spectrum
and can induce orbital currents. For continuous transitions, the instability is governed by the orbital susceptibility, connecting photon condensation to the magnetostatic Condon instability
\cite{Condon1966,AndolinaPRB20,GuerciPRB21}.
Related transitions have been explored in two-dimensional
electron gases with spin-orbit coupling
\cite{ManzanaresPRB22} and extended ladders
\cite{BacciconiSciPost23}. When the electronic energy gain outweighs the photon cost, a coherent cavity field develops without external driving.

At the molecular scale, Van Vleck paramagnetism
provides a route to orbital photon condensation
in ensembles of square plaquettes \cite{MercurioPRR24}.
Including Hubbard interactions and varying the
electronic filling modify both the threshold
and the order of the transition \cite{AndolinaCavity26}.
Here, we show that frustrated triangular Mott molecules
support photon condensation driven by an emergent
orbital degree of freedom: scalar spin chirality.
Virtual ring exchange \cite{MotrunichPRB06,KhomskiiPRB08} couples this orbital-current degree of freedom directly to cavity flux.
Invariant under spin rotations yet odd under time reversal, the scalar chirality is a key ingredient in frustrated magnets and chiral spin liquids \cite{Wen1989Chiral,KalmeyerLaughlin1987}.

Light-induced control of correlated magnetism has been
explored through laser-generated scalar-chirality
interactions \cite{KitamuraPRB17,ClaassenNC17},
photon-dependent exchange
\cite{PedrocchiPRB11,SentefPRR20,BostromNPJ23,Grunwald2026},
and cavity modification of magnetic frustration
and critical fluctuations
\cite{ChiocchettaNC21,WeberCP23}.
Vacuum fluctuations in gyrotropic cavities can also
stabilize chiral spin liquids \cite{WeiPRL25}.
Here, the orbital response of correlated spins
drives the cavity field to condense, spontaneously
breaking time-reversal symmetry and selecting
a finite scalar spin chirality.

{\it System and model Hamiltonian}. We consider a triangular Hubbard ring at half filling ($N_{\rm el}=3$), coupled to a single cavity mode as illustrated in Fig.~\ref{fig1}, and neglect
cavity losses. The cavity magnetic flux enters the electronic hopping through Peierls phases \cite{LiPRB20,DmytrukPRB21}, so that
the Hamiltonian is $H=H_{\rm el}+\omega_ca^\dagger a$
($\hbar=1$), with
\begin{align}
H_{\rm el}
={}&\sum_{j,\sigma}
\left(
t_j e^{ig_j(a^\dagger+a)}
c_{j\sigma}^\dagger c_{j+1\sigma}+\hc
\right)
+U\sum_j n_{j\uparrow}n_{j\downarrow}.
\label{eq:H_el}
\end{align}
Here $j=1,2,3$ labels sites cyclically around the triangle,
$t_j$ is the real hopping amplitude on bond $(j,j+1)$,
and $U$ is the uniform on-site Coulomb repulsion.
The operator $c^\dagger_{j\sigma}$ creates a spin-$\sigma$
electron on site $j$, while $a^\dagger$ creates a cavity
photon of frequency $\omega_c$.
The dimensionless coefficient $g_j$ specifies the
photon-induced Peierls phase on bond $(j,j+1)$ 
 \footnote{Gauge transformations in a quantized
light-matter system act on both electronic and photonic
degrees of freedom \cite{DmytrukPRB21,KozinPRR24}.
Redistributing the bond phases therefore requires
transforming the remaining Hamiltonian consistently
to preserve gauge equivalence.}.

{\it Effective spin-photon coupling}. In the Mott regime ($U\gg t_j$), the cavity couples to the low-energy spin degrees of freedom through virtual charge fluctuations. We project onto the singly occupied subspace $\mathcal P$, while virtual charge excitations in $\mathcal Q=1-\mathcal P$ are treated perturbatively through a Schrieffer-Wolff transformation (SWT) \cite{MacDonaldPRB88,SM}. Splitting the Hamiltonian as $H=H_0+V$, with
$H_0=U\sum_i n_{i\uparrow}n_{i\downarrow}+\omega_ca^\dagger a$
and $V$ the photon-dressed hopping, we obtain the effective
Hamiltonian to third order in $t_j/U$: 
\begin{align}
\!\!H_{\rm s-ph}
&\approx
\sum_{j}J_{j} {\bm S}_j\cdot{\bm S}_{j+1}
- K \sin\hat{\Phi}_{\rm cav}\;\chi+\omega_c a^\dagger a\,,
%\nonumber\\
%+\omega_ca^\dagger a-3K/4\cos\hat{\Phi}
\label{eq:H_sp}
\end{align}
where ${\bm S}_j=(1/2)\sum_{\sigma,\sigma'}\mathcal{P}c_{j\sigma}^\dagger({\boldsymbol \sigma})_{\sigma\sigma'}c_{j\sigma'}\mathcal{P}$ are the spin $1/2$ operators, $J_{j}=4|t_j|^2/U$, $K=24t_1t_2t_3/U^2$, $\hat{\Phi}_{\rm cav}=g(a^\dagger+a)$, and $g=\sum_{j=1}^3g_j$ \footnote{At the order considered here, the effective spin Hamiltonian depends only on the total operator-valued flux threading the triangle and is therefore insensitive to the particular distribution of Peierls phases among the bonds.}. The first term describes the antiferromagnetic superexchange, while the second couples the cavity field to the scalar spin chirality generated by three-site virtual hopping \cite{MotrunichPRB06,KhomskiiPRB08,TrifPRL08,TrifPRB10}
\begin{align}
\chi = {\bm S}_1 \cdot
\left(
{\bm S}_2 \times {\bm S}_3
\right)\,.
\end{align}
The scalar spin chirality is the low-energy representation of the orbital current generated by virtual electron hopping around the triangular plaquette. To make this connection explicit, consider the microscopic current operator on bond $(j,j+1)$,
\begin{equation}
\hat I_{j}
=
ie\,t_j
\sum_\sigma
\left(
c_{j\sigma}^\dagger c_{j+1\sigma}
-
c_{j+1\sigma}^\dagger c_{j\sigma}
\right).
\label{current}
\end{equation}
At zero flux, the SWT of the
current operator yields
$\mathcal{P} e^S\hat I_j e^{-S}\mathcal{P}=-I_0\chi$,
where $I_0=24e\,t_1t_2t_3/U^2$ and the sign follows
the bond-current convention in Eq.~\eqref{current}. The scalar spin chirality therefore represents the emergent orbital-current degree of freedom of the Mott triangle, providing the microscopic link between correlated electron motion and orbital magnetism~\cite{Taguchi_Science_01,KhomskiiPRB08}.

In deriving Eq.~\eqref{eq:H_sp}, we further assumed $U\gg\omega_c$, allowing photon-energy shifts in the virtual intermediate states to be neglected. The Supplemental Material (SM) \cite{SM} gives the corresponding expressions with photon-energy shifts retained in the virtual intermediate states. Because the cavity couples only through the Peierls phases, the global SU(2) spin symmetry is preserved. Consequently, the low-energy spectrum separates into a spin-$3/2$ quartet and two spin-$1/2$ doublets. The spin-$3/2$ sector is completely decoupled from the cavity, so the light-matter dynamics is entirely governed by the spin-$1/2$ manifold. Projecting onto this subspace therefore yields the minimal
low-energy Hamiltonian governing the coupled chirality-photon
dynamics,
\begin{align}
H_{\rm c-ph}
&=\frac{1}{2}\boldsymbol{\Delta}\cdot\boldsymbol C
-K'C_z\sin\hat{\Phi}_{\rm cav}
+\omega_ca^\dagger a\,,
\label{chir_phot}
\end{align}
where $\boldsymbol C=(C_x,C_y,C_z)$, with
\begin{align}
C_x&=\frac{2}{3}
(2{\bs S}_1\cdot{\bs S}_2
-{\bs S}_2\cdot{\bs S}_3
-{\bs S}_3\cdot{\bs S}_1)\,,
\nonumber\\
C_y&=\frac{2}{\sqrt{3}}
({\bs S}_3\cdot{\bs S}_1
-{\bs S}_2\cdot{\bs S}_3)\,,
\end{align}
and  $C_z=4\chi/\sqrt{3}$.  These operators act
as Pauli matrices in the $S=1/2$ subspace and vanish
in the $S=3/2$ sector. Here, $K'=\sqrt{3}K/4$ and
$\boldsymbol{\Delta}=[(2J_1-J_2-J_3)/2,
\sqrt{3}(J_3-J_2)/2,0]$, with $\Delta_z=0$ reflecting
the absence of a chirality bias at zero cavity flux.
Equation~\eqref{chir_phot} is the central result of this work:
in the Mott regime, the cavity couples directly to the
emergent scalar spin chirality, a many-body orbital degree
of freedom generated by strong electronic correlations.
%Unlike conventional cavity-QED systems, where the electric field couples to microscopic dipole moments, the magnetic cavity field couples through the Peierls phase.
\begin{figure}[t]
\centering
\includegraphics[width=\linewidth]{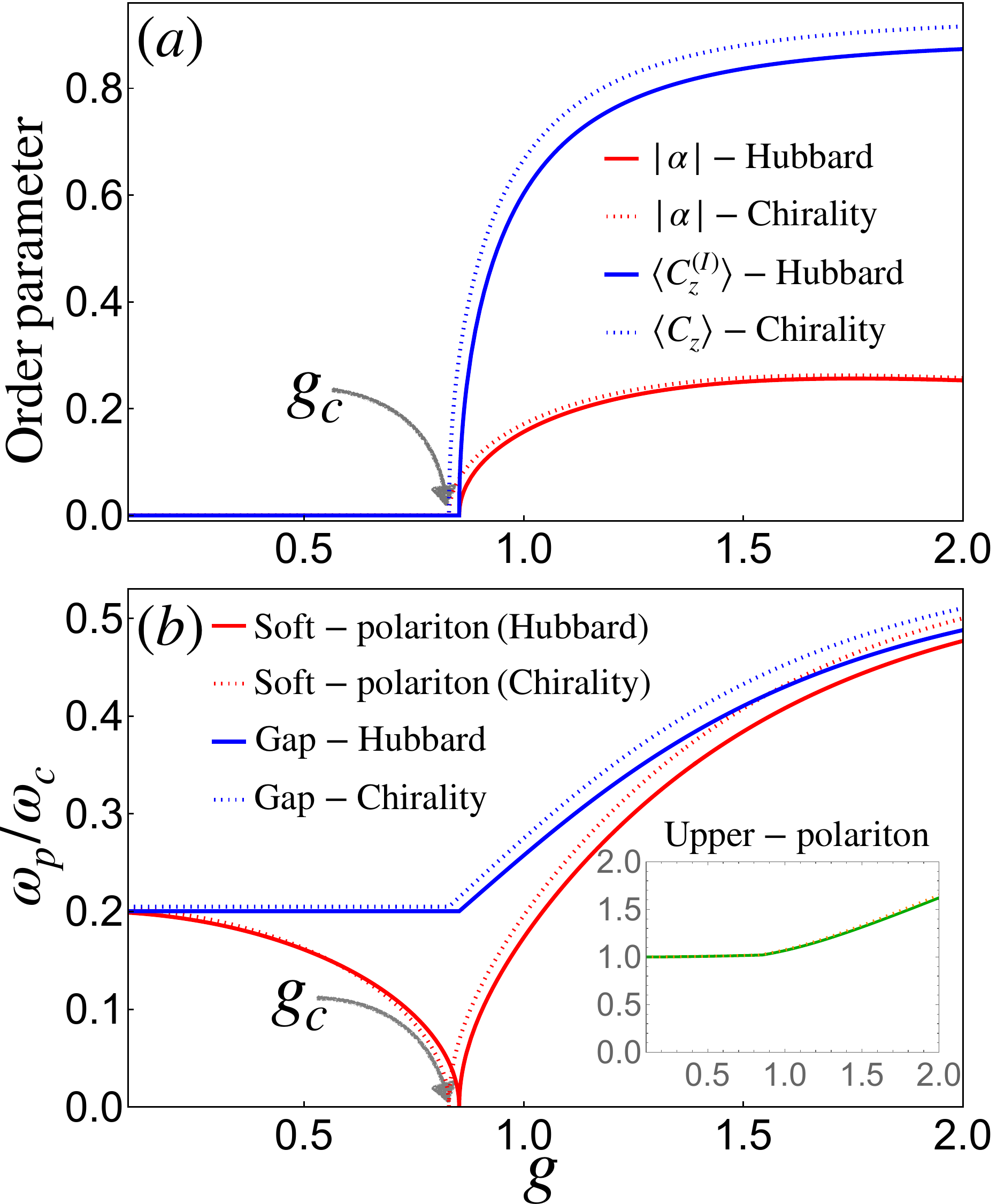}
\caption{Mean-field order parameters and collective excitation spectra of the full half-filled Hubbard model and the effective chirality model.
($a$) Mean-field order parameters as a function of the light-matter coupling $g$, showing the photon amplitude $|\alpha|$ and the chirality response for the full Hubbard model and the effective chirality model. The Hubbard chirality $\langle C_z^{(I)}\rangle$ is inferred from the circulating current using the third-order current–chirality relation and compared with the effective-model expectation value $\langle C_z\rangle$. In both cases, the order parameters turn on at their respective critical couplings, and the chirality model closely reproduces the Hubbard mean-field behavior.
($b$) Corresponding polaritonic excitations, $\omega_p$, with $p=\pm$, plotted in units of the cavity frequency $\omega_c$. The soft polariton frequency decreases toward zero at each model’s transition, while the local gap and the upper polariton frequency remain finite and are likewise well reproduced by the chirality model. Parameters: $t_1=t_2=1$, $t_3=1.05$, $U=20$ ($\Delta=0.0205$), and $\omega_c=0.1$.}
\label{fig2}
\end{figure}

{\it Mean-field description of the photons}. We now consider $N$ identical triangles coupled to a common
cavity mode, with the per-triangle coupling scaled as
$g/\sqrt N$. For the triangle ensemble, the Hamiltonian is
\begin{equation}
H_N=\omega_c a^\dagger a+
\sum_{n=1}^{N}\left[
\frac12\boldsymbol{\Delta}\cdot\boldsymbol C_{n}
-K'C_{z,n}\sin\!\left(\frac{\hat\Phi_{\rm cav}}{\sqrt N}\right)
\right]\,.
\end{equation}
At zero temperature, we approximate the ground state by $|\Psi_{\rm tot}\rangle=|\psi_c\rangle\otimes|\sqrt N\,\alpha\rangle$,
where $\alpha=\langle a\rangle/\sqrt N$ is real, following the coherent-state mean-field treatment
of orbital photon condensation
\cite{MercurioPRR24,AndolinaCavity26}.
The coherent-state average of the sinusoidal coupling is
$e^{-g^2/(2N)}\sin(2g\alpha)$. Its prefactor tends to unity as $N\to\infty$, giving the ground-state energy per triangle
\begin{equation}
E_{GS}(\alpha)
=\omega_c\alpha^2
-\sqrt{(\Delta/2)^2+K'^2\sin^2(2g\alpha)}\,,
\label{eq:energy_functional}
\end{equation}
where $\Delta=|\boldsymbol{\Delta}|$ is the chirality splitting. Expanding the energy around the normal phase, $
E_{GS}(\alpha)=E_{GS}(0)+A(g)\alpha^2+\cdots$,
with
$A(g)=\omega_c-4g^2K'^2/\Delta$
shows that the quadratic coefficient changes sign at
$g_c=\frac{\sqrt{\omega_c\Delta}}{2|K'|}$, 
marking the onset of a superradiant transition accompanied by spontaneous scalar chirality, $\langle C_z\rangle\neq0$, and a macroscopic cavity displacement. The transition is therefore driven by the collective ordering of an emergent chirality degree of freedom rather than by conventional spin polarization. %The finite displacement also renormalizes the chirality gap and light-matter coupling, giving rise to the chirality polaritons discussed below.

To benchmark the effective theory, Fig.~\ref{fig2}(a)
compares independently determined mean-field solutions
of the full three-electron Hubbard model and the
effective chirality model. The Hubbard chirality signal
is inferred from the circulating current using the
third-order current-chirality relation. Their close agreement supports the low-energy description of the orbital response in the Mott regime. At fixed $K'$, the dependence $g_c\propto\sqrt{\Delta}$
enables \emph{in situ} electrical control of the
superradiant threshold through Stark tuning of
$\Delta(\bs E)$
\cite{TrifPRL08,TrifPRB10,Islam2010}. This separates the control mechanism from the light-matter interaction: an applied electric field tunes the molecular chirality splitting, while the cavity couples through orbital magnetic flux, without requiring an external magnetic bias. In molecular magnets with triangular spin units,
such as Cu$_3$ and V$_{15}$, exchange inequivalence
produces a chirality splitting that can be modified
by an applied electric field \cite{Islam2010,NossaPRB23}.
Recently, such electric-field tuning of the exchange interaction has been demonstrated in a Cu$_3$ spin triangle \cite{Cini2025}.

{\it Chirality polaritons}. We now investigate collective light–matter excitations around the mean-field solution, neglecting spin–orbit coupling. Writing $a=\sqrt N\,\alpha+d$ and describing
the chirality fluctuations through a Holstein–Primakoff representation \cite{HolsteinPrimakoff40,EmaryPRE03, BaksicPRL14}
gives the quadratic Hamiltonian of the cavity-active
bright sector \cite{SM}
\begin{equation}
H_{\rm pol}=\omega_c d^\dagger d+A_\alpha(d^\dagger+d)^2+\Delta_{\alpha}b^\dagger b+G_{\alpha}(d+d^\dagger)(b+b^\dagger)\,,
\label{eq:Hfl}
\end{equation}
where only the fully symmetric bright mode $b=\frac1{\sqrt N}\sum_n b_n$ hybridizes with the photon field, where $b_n$ is the annihilation operator for a local chirality excitation on triangle $n$. This collective bright mode is a direct consequence of the permutation symmetry of the molecular triangles and is analogous to the Dicke model, whereas the remaining $N-1$ orthogonal modes remain dark. The equilibrium cavity displacement determines the effective chirality gap $\Delta_\alpha$, the light-matter coupling
$G_\alpha$, and the cavity curvature coefficient $A_\alpha$:
\begin{align}
\Delta_{\alpha}&=\sqrt{\Delta^2+[2K'\sin(2g\alpha)]^2}\,,\nonumber\\
G_\alpha
&=-\frac{gK'\Delta\cos(2g\alpha)}{\Delta_\alpha}\,,\\
A_\alpha
&=\frac{g^2K'^2\sin^2(2g\alpha)}{\Delta_\alpha}\,\nonumber,
\end{align}
where  $A_\alpha$ arises from the curvature of the sinusoidal
Peierls coupling.  Diagonalization yields the upper ($+$) and lower ($-$) polariton branches:
\begin{equation}
\omega_\pm^2=\frac{\widetilde{\omega}_c^2+\Delta_\alpha^2}{2}\pm\frac{1}{2}\sqrt{\left(\widetilde{\omega}_c^2-\Delta_\alpha^2\right)^2+16G_{\alpha}^2\omega_c\Delta_\alpha}\,,
\label{eq:polariton_frequencies}
\end{equation}
with $\widetilde{\omega}_c=\sqrt{\omega_c(\omega_c+4A_\alpha)}$.  As shown in Fig.~\ref{fig2}($b$), the lower polariton
branch softens and vanishes at the continuous
superradiant transition, providing a direct spectroscopic
signature of the collective instability. 
 
Remarkably, the resulting polariton spectrum is in excellent quantitative agreement with that
obtained from collective fluctuations around the mean-field
solution of the full microscopic Hubbard model, demonstrating that the effective chirality theory faithfully captures the low-energy collective physics of the microscopic model. 

\begin{figure}[t]
\centering
\includegraphics[width=\linewidth]{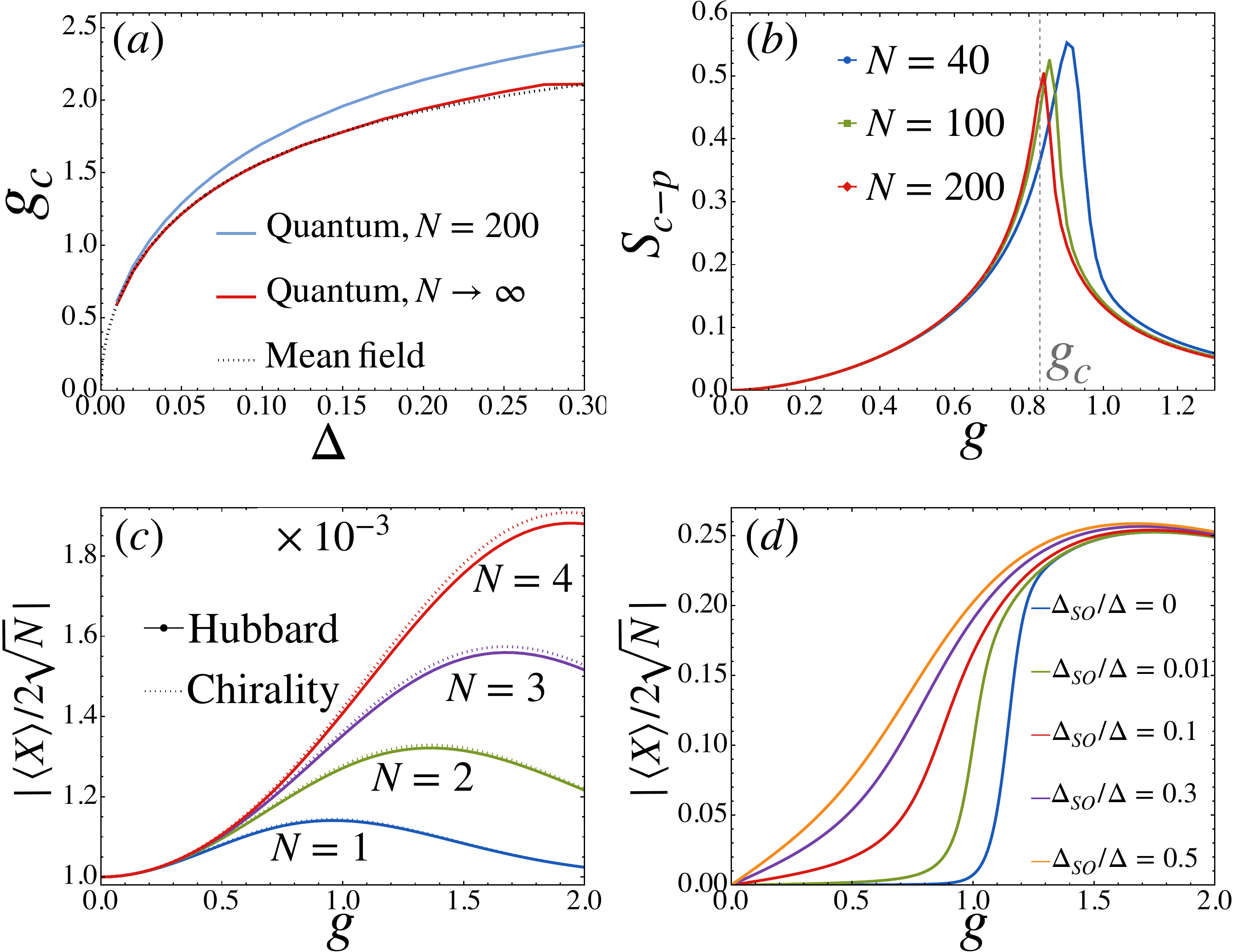}
\caption{
Quantum characterization of cavity-induced superradiance.
($a$) Critical coupling versus chirality splitting, obtained
by varying $t_3=\sqrt{t^2+U\Delta/4}$ at fixed
$t_1=t_2=t=1$, $U=20$, and $\omega_c=0.1$.
The mean-field curve uses the corresponding
$K'(\Delta)=6\sqrt{3}\,t^2t_3/U^2$.
Quantum results are shown for $N=200$, together
with extrapolated thermodynamic estimates \cite{SM}.
($b$) Chirality-photon entanglement entropy $S_{c-p}$
for $N=40$, $100$, and $200$ triangles.
Its maximum sharpens and approaches the mean-field
critical coupling $g_c$ (vertical dashed line)
as $N$ increases. A selecting field $h\sqrt N(a+a^\dagger)$, with $h=10^{-4}$, is included. ($c$) Photon displacement for the microscopic Hubbard model (solid lines) and effective chirality model (dashed lines), for $N=1,\ldots,4$ triangles, demonstrating their close agreement in the Mott regime. ($d$) Photon displacement for $N=20$ in a fixed,
fully polarized spin sector for several values of
$\Delta_{\rm SO}/\Delta$, with selecting-field
magnitude $h=10^{-5}$.
SOI acts as a longitudinal chirality bias.
The curves illustrate finite-size crossovers
rather than thermodynamic critical thresholds. Here $X=a+a^\dagger$, and the per-triangle cavity coupling is $g/\sqrt N$. In ($c$), $U=30$ and $H_{\rm bias}=h\sqrt N X$, while the remaining microscopic parameters are as in Fig.~\ref{fig2}.
}
\label{fig3}
\end{figure}

{\it Exact diagonalization}. To benchmark the theory beyond
mean field, we perform exact diagonalization (ED) of the
effective chirality-photon Hamiltonian and the full microscopic
electron-photon model. For the effective model, we work in the
permutation-symmetric sector with up to $N=240$ triangles,
supplemented by calculations through $N=400$ at two
representative parameter points. We locate the finite-size
pseudocritical coupling from the maximum slope of the
parity-even photon amplitude
$\alpha_{\rm fl}=\sqrt{\max[0,(\langle X^2\rangle-1)/(4N)]}$,
where $X=a+a^\dagger$ \cite{EmaryPRE03}.
Figure~\ref{fig3}($a$) compares these couplings and their
extrapolated estimates with the mean-field phase boundary.
Fits based on the leading $N^{-2/3}$ scaling
\cite{Vidal_2006} closely reproduce the mean-field prediction.
Numerical convergence and extrapolation sensitivity are
discussed in the SM \cite{SM}.

As an independent diagnostic, Fig.~\ref{fig3}($b$) shows
the chirality-photon entanglement entropy
\cite{VidalPRL03,LambertPRL04},
\begin{equation}
S_{c-p}=-{\rm Tr}\!\left(\rho_{\rm ph}\ln\rho_{\rm ph}\right),
\end{equation}
where $\rho_{\rm ph}={\rm Tr}_{\chi}|\Psi_0\rangle
\langle\Psi_0|$ is the reduced photonic density matrix.
The ground state $|\Psi_0\rangle$ is evaluated with a weak
selecting field $H_{\rm bias}=h\sqrt N X$, with $h=10^{-4}$.
The entropy peak sharpens and moves toward the mean-field
critical coupling with increasing system size, providing
an additional quantum signature of the collective instability.

Finally, Fig.~\ref{fig3}($c$) compares the photon displacement
obtained by ED of up to $N=4$ half-filled Hubbard triangles
($N_{\rm el}=3$ per triangle) with that of the effective
chirality model. Their close agreement shows that the
three-spin ring-exchange interaction captures the low-energy
light-matter coupling in the Mott regime, with only small
quantitative corrections from charge fluctuations.

{\it Effects of spin-orbit interaction}. We now examine the robustness of the superradiant transition in the presence of  the spin-orbit interaction (SOI). The leading effect of SOI is to couple the emergent scalar chirality to the local spin polarization \cite{TrifPRL08, TrifPRB10}, pertaining to the Hamiltonian $H_{\rm SO} = \Delta_{\rm SO} \sum_{n=1}^N C_{z,n} S_{z,n}$, while neglecting SOI-dependent corrections to the orbital light-matter coupling.  Within this projected model, each $S_{z,n}$ is conserved,
and the Hamiltonian separates into sectors of fixed
molecular spin projections. For nonzero SOI and a nonzero equilibrium cavity displacement, the unconstrained ground state selects a fully polarized spin sector. Within a fixed spin sector, the SOI acts as a longitudinal bias on chirality. However, when both spin sectors are allowed, the two displaced branches remain symmetry-related. At zero temperature, minimizing the mean-field energy over both spin sectors produces a cusp at zero cavity
displacement, so that any nonzero light-matter coupling
favors symmetry-related displaced minima.  Figure~\ref{fig3}($d$) shows the corresponding displacement response within a fixed polarized sector. By contrast, fixing the total magnetization to zero gives equal populations of the two spin sectors, whose linear chirality biases cancel. The ground-state energy is then analytic around zero displacement, and the normal-state instability occurs at
\begin{equation}
g_c(\Delta_{\rm SO})
=
g_c(0)
\left[
1+\left(\frac{\Delta_{\rm SO}}{\Delta}\right)^2
\right]^{3/4}.
\label{eq:gc_soi}
\end{equation}
This threshold increases with SOI because the spin-chirality
coupling reduces the susceptibility to cavity-induced
chirality ordering. It coincides with the transition
coupling when the transition is continuous.

{\it Estimates and candidate molecular systems}. We focus on superconducting LC resonators because the proposed mechanism
relies on the orbital coupling to the magnetic flux generated by the cavity mode, rather than on the electric-dipole interaction dominant in conventional cavity QED. The zero-point current produces a flux $\Phi_{{\rm zpf},i}$ through each molecular triangle and hence a dimensionless Peierls coupling $g_i=2\pi\Phi_{{\rm zpf},i}/\Phi_e$, where
$\Phi_e=h/e$ is the electronic flux quantum. For a conventional circular loop and an ensemble of
$N\sim10^{12}$ molecules, we obtain the conservative
estimate \cite{SM}
\begin{equation}
g=\sqrt{\sum_i g_i^2}\sim10^{-6}\,.
\end{equation}
The corresponding collective energy scale is
$\Delta_{\rm cav}\sim |K'|g$, giving
$\Delta_{\rm cav}/h\sim10$--$100~{\rm kHz}$
for an illustrative range
$|K'|/h\sim10$--$100~{\rm GHz}$, bracketed by
the molecular estimates below.
Increasing this scale motivates resonator geometries
that concentrate the zero-point magnetic field within
the molecular sample, such as narrow inductive
constrictions \cite{GimenoACSNano20}. %Experimentally robust couplings therefore require resonator geometries that concentrate the zero-point magnetic flux within the molecular sample, such as narrow inductive constrictions or loop-gap structures.

As a quantitative benchmark, approximate Hubbard
parameters reported for Cu$_3$ molecular triangles
\cite{NossaPRB23}, $t\simeq50~{\rm meV}$ and
$U\simeq9.06~{\rm eV}$, give
$|K'|/h\simeq4~{\rm GHz}$ using
$|K'|=6\sqrt{3}|t|^3/U^2$.
For V$_3$ molecular triangles \cite{NossaPRB23},
$t\simeq53~{\rm meV}$ and $U\simeq1~{\rm eV}$
yield $|K'|/h\simeq370~{\rm GHz}$.
The enhancement primarily reflects the smaller
effective Coulomb repulsion, illustrating the potential
of molecular design to strengthen the orbital coupling.
These values remain indicative: for V$_3$, the same
one-band parameters give $J=4t^2/U\simeq11~{\rm meV}$,
substantially exceeding the reported DFT exchange
of $1.3~{\rm meV}$. A quantitative prediction, therefore, requires a
consistent microscopic parametrization.
More broadly, the scale $t_1t_2t_3/U^2$ provides
a microscopic guide to enhancing the orbital
chirality-photon coupling.

{\it Conclusions and outlook}. We have shown that
orbital cavity coupling can drive equilibrium photon
condensation and spontaneous scalar spin chirality
in triangular Mott molecules.
Our microscopic derivation and numerical benchmarks
establish an effective chirality-photon description,
with collective polariton excitations and an
electrically tunable condensation threshold.

A natural extension is to evaluate the finite-frequency
exchange and chirality couplings derived in the SM,
following the approach developed for cavity-engineered
exchange \cite{SentefPRR20}. This would reveal how
photon-assisted ring exchange modifies chirality and
orbital currents beyond the low-frequency limit.
Beyond molecular magnets, gate-defined quantum-dot
triangles offer another setting for exploring
chirality-photon coupling, motivated by the recent
demonstration of simultaneous control of all three
exchange interactions \cite{BrozPRXQuantum26}.
Coupling a single triangle to the electric field of
a resonator could enable manipulation of its chirality
degree of freedom through electrically modulated
exchange. Progress in controlling two-dimensional
quantum-dot arrays \cite{JohnNatCommun25} also
motivates extending this approach to several
triangular units coupled to a common resonator.
More broadly, our work opens a route to manipulating
frustrated magnetism through cavity control of
emergent orbital degrees of freedom.

{\it Acknowledgments}. We thank V. K. Kozin for very useful discussions. M. T. acknowledges support from
 the National Science Centre (Poland) OPUS Grant No.~2021/42/B/ST3/04475, the Foundation for
 Polish Science project ``MagTop'' (No.~FENG.02.01-IP.05-0028/23) cofinanced by the European Union from
 the funds of Priority 2 of the European Funds for a Smart Economy Program 2021--2027 (FENG), and by
 the NAWA Bekker Grant No.~BPN/BEK/2024/1/00310 (Poland). D. S. acknowledges funding provided by
 the Institute of Physics Belgrade, through a grant from the Ministry of Science, Technological Development
 and Innovation of the Republic of Serbia.

{\it Data availability}. The data and exact-diagonalization scripts supporting this work are available from the authors upon reasonable request.

\clearpage
\onecolumngrid

\begin{center}
{\large\bfseries
Supplemental Material for\\[2pt]
``Photon condensation driven by scalar spin chirality''
\par}

\vspace{0.4cm}

Dimitrije Stepanenko$^{1}$,
Daniel Loss$^{2,3,4}$,
and Mircea Trif$^{5}$

\vspace{0.25cm}

{\small\itshape
$^{1}$Nanostructured Matter Laboratory,
Center for Solid State Physics and New Materials,\\
Institute of Physics Belgrade, University of Belgrade,
Pregrevica 118, 11080 Belgrade, Serbia\\[3pt]

$^{2}$Physics Department, King Fahd University of
Petroleum and Minerals, 31261, Dhahran, Saudi Arabia\\[3pt]

$^{3}$Quantum Center and RDIA Chair in Quantum Computing,
KFUPM, Dhahran\\[3pt]

$^{4}$Department of Physics, University of Basel,
Klingelbergstrasse 82, 4056 Basel, Switzerland\\[3pt]

$^{5}$International Research Centre MagTop,
Institute of Physics, Polish Academy of Sciences,\\
Aleja Lotnikow 32/46, PL-02668 Warsaw, Poland
\par}
\end{center}

\vspace{0.3cm}

This Supplemental Material presents the microscopic
derivation of the chirality-photon Hamiltonian,
retaining photon-energy shifts in the virtual
intermediate states. We derive the collective excitation
spectra for the effective chirality and microscopic
Hubbard models, document numerical convergence and
finite-size extrapolation, analyze the effects of
spin-orbit coupling and finite temperature, and
estimate the orbital light-matter coupling for
molecular ensembles.

\section{Schrieffer-Wolff derivation of three-site ring exchange}
\label{sec:SM_SW}

We start by writing  the total Hamiltonian as $H=H_U+H_t+H_p$, with
\[
H_U=U\sum_j n_{j\uparrow}n_{j\downarrow}\,,\qquad 
H_t=\sum_{j,\sigma}
\left[t_j e^{ig_j(a^\dagger+a)}
c^\dagger_{j\sigma}c_{j+1\sigma}+{\rm H.c.}\right]\,,\qquad H_p=\omega_c a^\dagger a\,,
\]
where we assume $t_j$ are real. To systematically account for photon-assisted virtual processes, it is convenient to work in a frame rotating at the cavity frequency. In this representation, the explicit photon Hamiltonian is removed, while the Peierls phases become explicitly time periodic. This allows us to employ a Floquet-SWT expansion \cite{SentefPRR20}. Specifically, we perform a unitary transformation $\mathcal{U}(\tau)=e^{i\omega_ca^\dagger a\tau}$ on the total triangle-cavity Hamiltonian, which removes $H_p$ and leads to the replacement $a\rightarrow ae^{-i\omega_c\tau}$. The transformed Hamiltonian becomes
\begin{align}
H_{\rm rot}(\tau)-i\partial_\tau
&=\mathcal U(\tau)[H-i\partial_\tau]\mathcal U^\dagger(\tau),
\qquad
    H_{\rm rot}(\tau)=H_U+H_t(\tau)\,,\\
    H_t(\tau)&=\sum_{j,\sigma}
\left[t_j e^{ig_jA(\tau)} c^\dagger_{j\sigma}c_{j+1\sigma}+{\rm H.c.}\right]\,,
\end{align}
where $A(\tau)=(a^\dagger e^{i\omega_c\tau}+ae^{-i\omega_c\tau})$. Note that this transformation renders the Hamiltonian periodic, $H_{\rm rot}(\tau+T_c)=H_{\rm rot}(\tau)$, with $T_c=2\pi/\omega_c$. This periodicity is a consequence of the rotating-frame representation of the quantized cavity mode and does
not introduce an external drive.  At half filling, the Hilbert space is partitioned into the singly occupied sector, with projector $P$, and its complement $Q=1-P$. The projectors act on the electronic Hilbert space at fixed particle number $N_{\rm el}=3$ and are understood to include the identity in photon space. For the three-site ring, the latter contains exactly
one doublon and one holon, with interaction energy $U$, so that
$PH_UP=0$ and $QH_UQ=UQ$. A single hopping process acting on
the singly occupied sector creates a doublon-holon pair,
implying $PH_tP=0$, whereas hopping within $Q$ can move the
doublon or holon without changing the interaction energy.
We therefore introduce the off-diagonal and diagonal perturbations:
\begin{align}
V_{od}(\tau)&=PH_t(\tau)Q+QH_t(\tau)P\,,\nonumber\\
V_{d}(\tau)&=\underbrace{PH_t(\tau)P}_{=0}+QH_t(\tau)Q\,.
\end{align}

Next, we use a time-dependent SWT $W(\tau)=e^{S(\tau)}$ that block-diagonalizes $H$ order by order. We choose an anti-Hermitian, block-off-diagonal generator, $S^\dagger=-S$, with $PSP=QSQ=0$.
We use $O_{00}=POP$, $O_{11}=QOQ$,
$O_{10}=QOP$, and $O_{01}=POQ$, so that 
\begin{equation}
\widetilde H_{\rm rot}(\tau)-i\partial_\tau
=W(\tau)[H_{\rm rot}(\tau)-i\partial_\tau]W^\dagger(\tau).
\end{equation}
We introduce a dimensionless bookkeeping parameter $\lambda$ by writing $H_{\rm rot}=H_U+\lambda(V_{od}+V_d)$ and expand the generator as $S(\tau)=\lambda S_1(\tau)+\lambda^2S_2(\tau)$.
Each power of $\lambda$ tracks the corresponding order in the
hopping amplitudes; we set $\lambda=1$ at the end.
Up to third order in $\lambda$, we obtain:
\begin{align}
   \widetilde{H}_{\rm rot}(\tau)&=H_{U}+\lambda(V_{od}+V_{d})+\lambda[S_1,H_U]+i\lambda\dot{S}_1\nonumber\\
   &+\lambda^2[S_1,V_{od}+V_d]+\lambda^2[S_2,H_U]+\frac{\lambda^2}{2}[S_1,[S_1,H_U]]+i\frac{\lambda^2}{2}(S_1\dot{S}_1-\dot{S}_1S_1)+i\lambda^2\dot{S}_2\nonumber\\
   &+\lambda^3[S_2,V_{od}+V_d]+\frac{\lambda^3}{2}[S_2,[S_1,H_U]]+\frac{\lambda^3}{2}[S_1,[S_2,H_U]]+\frac{\lambda^3}{2}[S_1,[S_1,V_{od}+V_d]]+\frac{\lambda^3}{6}[S_1,[S_1,[S_1,H_U]]]\nonumber\\
   &+\frac{i\lambda^3}{6}[S_1,[S_1,\dot{S}_1]]
   +\frac{i\lambda^3}{2}
   \left([S_1,\dot{S}_2]+[S_2,\dot{S}_1]\right)
   +O(\lambda^4)\,.
\end{align}
We truncate the generator at
$S(\tau)=\lambda S_1(\tau)+\lambda^2S_2(\tau)$,
which suffices to obtain the projected low-energy Hamiltonian
through third order in the hopping. A third-order block-off-diagonal generator contributes only through
$[S_3,H_U]+i\dot S_3$, whose projection onto the singly occupied
sector vanishes. Next, we find $S_1$ and $S_2$ order by order by eliminating the transition between the two blocks up to second order:
\begin{align}
    [S_1,H_U]+V_{od}+i\dot{S}_1=0\,,\\
    [S_2,H_U]+[S_1,V_d]+i\dot{S}_2=0\,,
\end{align}
which, in turn, leads to the following effective low-energy Hamiltonian:
\begin{align}
H_{\rm eff}(\tau)
\equiv P\widetilde H_{\rm rot}(\tau)P
&=\frac{\lambda^2}{2}P[S_1(\tau),V_{od}(\tau)]P
+\frac{\lambda^3}{2}P[S_2(\tau),V_{od}(\tau)]P
+\mathcal O(\lambda^4).
\label{eq:Heff-orders}
\end{align}
Here $PH_UP=0$. Although the truncated transformation leaves
third-order off-diagonal terms, these vanish upon projection
and do not affect $H_{\rm eff}$ at this order. The generators $S_1$ and $S_2$ are conveniently obtained in Fourier space, since all operators inherit the cavity periodicity:
\begin{align}
    O(\tau)&=\sum_nO^{(n)}e^{-in\omega_c\tau}\,,\\
    O^{(n)}&=\frac{1}{T_c}\int_0^{T_c}d\tau O(\tau) e^{in\omega_c\tau}\,,
\end{align}
so that 
\begin{align}
V_{od}^{(n)}&=\sum_{j,\sigma}t_jB_{g_j}^{(n)}\left[\underbrace{c_{j\sigma}^\dagger n_{j\bar{\sigma}} c_{j+1\sigma}(1-n_{j+1\bar{\sigma}})}_{\propto V_{od,10}}+\underbrace{c_{j\sigma}^\dagger(1-n_{j\bar{\sigma}}) c_{j+1\sigma}n_{j+1\bar{\sigma}}}_{\propto V_{od,01}}\right]\nonumber\\
    &+\sum_{j,\sigma}t_jB_{-g_j}^{(n)}\left[\underbrace{c_{j+1\sigma}^\dagger n_{j+1\bar{\sigma}}c_{j\sigma}(1-n_{j\bar{\sigma}})}_{\propto V_{od,10}}+\underbrace{c_{j+1\sigma}^\dagger(1-n_{j+1\bar{\sigma}}) c_{j\sigma}n_{j\bar{\sigma}}}_{\propto V_{od,01}}\right]\,,\\
V_{d}^{(n)}&=\sum_{j,\sigma}t_j\left[B_{g_j}^{(n)}c_{j\sigma}^\dagger(1- n_{j\bar{\sigma}})c_{j+1\sigma}(1-n_{j+1\bar{\sigma}})+B_{-g_j}^{(n)}c_{j+1\sigma}^\dagger(1-n_{j+1\bar{\sigma}}) c_{j\sigma}(1-n_{j\bar{\sigma}})\right]\nonumber\\
&+\sum_{j,\sigma}t_j\left[B_{g_j}^{(n)}c_{j\sigma}^\dagger n_{j\bar{\sigma}}c_{j+1\sigma}n_{j+1\bar{\sigma}}+B_{-g_j}^{(n)}c_{j+1\sigma}^\dagger n_{j+1\bar{\sigma}} c_{j\sigma}n_{j\bar{\sigma}}\right]\\
    S_{1}^{(n)}&=\frac{V_{od,10}^{(n)}}{U-n\omega_c}-\frac{V_{od,01}^{(n)}}{U+n\omega_c}=\sum_{j,\sigma}t_j\left[(c_{j\sigma}^\dagger c_{j+1\sigma})_{10}B_{g_j}^{(n)}+(c_{j+1\sigma}^\dagger c_{j\sigma})_{10}B_{-g_j}^{(n)}\right]\frac{1}{U-n\omega_c}\nonumber\\
    &-\sum_{j,\sigma}t_j\left[(c_{j\sigma}^\dagger c_{j+1\sigma})_{01}B_{g_j}^{(n)}+(c_{j+1\sigma}^\dagger c_{j\sigma})_{01}B_{-g_j}^{(n)}\right]\frac{1}{U+n\omega_c}\nonumber\\
    &=\sum_{j,\sigma}t_j\left[c_{j\sigma}^\dagger n_{j\bar{\sigma}} c_{j+1\sigma}(1-n_{j+1\bar{\sigma}})B_{g_j}^{(n)}+c_{j+1\sigma}^\dagger n_{j+1\bar{\sigma}}c_{j\sigma}(1-n_{j\bar{\sigma}})B_{-g_j}^{(n)}\right]\frac{1}{U-n\omega_c}\nonumber\\
    &-\sum_{j,\sigma}t_j\left[c_{j\sigma}^\dagger(1-n_{j\bar{\sigma}}) c_{j+1\sigma}n_{j+1\bar{\sigma}}B_{g_j}^{(n)}+c_{j+1\sigma}^\dagger(1-n_{j+1\bar{\sigma}}) c_{j\sigma}n_{j\bar{\sigma}}B_{-g_j}^{(n)}\right]\frac{1}{U+n\omega_c}\,,
\end{align}
where 
\begin{align}
    B_{g_j}^{(n)}&=\frac{1}{T_c}\int_0^{T_c}d\tau e^{ig_jA(\tau)}e^{in\omega_c\tau}=\frac{e^{-g_j^2/2}}{T_c}\int_0^{T_c}d\tau e^{ig_ja^\dagger e^{i\omega_c\tau}}e^{ig_ja e^{-i\omega_c\tau}}e^{in\omega_c\tau}\nonumber\\
    &=e^{-g_j^2/2}
\sum_{k=\max(0,-n)}^\infty
(ig_j)^{n+2k}
\frac{(a^\dagger)^k a^{n+k}}{k!\,(n+k)!}\,.
\end{align}
The Fourier component $B_g^{(n)}$ changes the photon
number by $-n$, so that positive $n$ corresponds
to photon annihilation. We can also extract the time-dependent transformation:
\begin{align}
    S_1(\tau)&=\sum_{j,\sigma}t_j\left[c_{j\sigma}^\dagger n_{j\bar{\sigma}} c_{j+1\sigma}(1-n_{j+1\bar{\sigma}})B_{g_j}^{-}(\tau)+c_{j+1\sigma}^\dagger n_{j+1\bar{\sigma}}c_{j\sigma}(1-n_{j\bar{\sigma}})B_{-g_j}^{-}(\tau)\right]\nonumber\\
    &-\sum_{j,\sigma}t_j\left[c_{j\sigma}^\dagger(1-n_{j\bar{\sigma}}) c_{j+1\sigma}n_{j+1\bar{\sigma}}B_{g_j}^{+}(\tau)+c_{j+1\sigma}^\dagger(1-n_{j+1\bar{\sigma}}) c_{j\sigma}n_{j\bar{\sigma}}B_{-g_j}^{+}(\tau)\right]\,,
\end{align}
where
\begin{align}
   B_{g_j}^{\pm}(\tau)=\sum_nB_{g_j}^{(n)}\frac{e^{-in\omega_c\tau}}{U\pm n\omega_c}\,. 
\end{align}
This allows us to evaluate the second-order correction to the Hamiltonian in the singly occupied block:
\begin{align}
\Delta H_2&=\frac{1}{2}[S_1(\tau),V_{od}(\tau)]_{00}=\sum_{n}\Delta H_2^ne^{-in\omega_c\tau}\,,\\
\Delta H_2^n&=-\frac{1}{2}\sum_{m,l}\delta_{m+l,n}\left(\frac{V_{od,01}^{(m)}V_{od,10}^{(l)}}{U+m\omega_c}+\frac{V_{od,01}^{(m)}V_{od,10}^{(l)}}{U-l\omega_c}\right)\nonumber\\
&=-\frac{1}{2}\sum_{m,l}\sum_{j,\sigma,\sigma'}\delta_{m+l,n}t_j^2\bigg[B_{g_j}^{(m)}B_{-g_j}^{(l)}c_{j\sigma}^\dagger(1-n_{j\bar{\sigma}})c_{j+1\sigma}n_{j+1\bar{\sigma}}c^{\dagger}_{j+1\sigma'}n_{j+1\bar{\sigma}'}c_{j\sigma'}(1-n_{j\bar{\sigma}'})\nonumber\\
&+B_{-g_j}^{(m)}B_{g_j}^{(l)}c_{j+1\sigma}^\dagger (1-n_{j+1\bar{\sigma}})c_{j\sigma}n_{j\bar{\sigma}}c^{\dagger}_{j\sigma'}n_{j\bar{\sigma}'}c_{j+1\sigma'}(1-n_{j+1\bar{\sigma}'})\bigg]\left(\frac{1}{U+m\omega_c}+\frac{1}{U-l\omega_c}\right)\nonumber\\
&=-\frac{1}{2}\sum_{m,l}\sum_{j}\delta_{m+l,n}t_j^2\bigg(B_{g_j}^{(m)}B_{-g_j}^{(l)}+B_{-g_j}^{(m)}B_{g_j}^{(l)}\bigg)(1/2-2S_j^zS_{j+1}^z-S_j^+S_{j+1}^--S_j^-S_{j+1}^+)\left(\frac{1}{U+m\omega_c}+\frac{1}{U-l\omega_c}\right)\nonumber\\
&=\sum_{m,l}\sum_{j}\delta_{m+l,n}t_j^2\bigg(B_{g_j}^{(m)}B_{-g_j}^{(l)}+B_{-g_j}^{(m)}B_{g_j}^{(l)}\bigg)\underbrace{({\bs S}_j\cdot{\bs S}_{j+1}-1/4)}_{-\mathcal{P}_{jj+1}}\left(\frac{1}{U+m\omega_c}+\frac{1}{U-l\omega_c}\right)\,,
\end{align}
where $\mathcal{P}_{jj+1}$ is the projector onto the singlet state on the bond $j$. This expression represents the photon-dressed superexchange interaction. In general, the exchange acquires an operator-valued dependence on the cavity field through the Fourier coefficients $B_g^{(n)}$ and the photon-dependent energy denominators. In the low-frequency limit $U\gg\omega_c$, the photon energies may be neglected in the denominators, recovering the conventional Heisenberg exchange
\begin{align}
    \Delta H_2=\sum_{j}J_j({\bs S}_{j}\cdot{\bs S}_{j+1}-1/4)\,,
\end{align}
where $J_j=4t_j^2/U$ represents the effective strength of the nearest-neighbor Heisenberg exchange interaction. Put differently, in this limit the dynamics reduces to particles tunneling back and forth between the two sites, so that the Peierls phase is fully canceled in the effective process. 

The third-order contribution follows analogously, but is qualitatively richer because the cyclic hopping process involves two successive virtual intermediate states containing a doublon–holon pair. Consequently, the effective chirality interaction depends on products of photon-dressed denominators, providing the natural extension of cavity-engineered superexchange to emergent scalar spin chirality.  We can extract the $n$-th Fourier component of $S_2(\tau)$:
\begin{align}
    S_{2}^{(n)}=\frac{[S_1(\tau),V_d(\tau)]_{10}^{(n)}}{U-n\omega_c}-\frac{[S_1(\tau),V_d(\tau)]_{01}^{(n)}}{U+n\omega_c}\,.
\end{align}
To proceed, we must first evaluate the $n$-th Fourier component of the commutator $[S_1(\tau),V_d(\tau)]$:
\begin{align}
[S_1(\tau),V_d(\tau)]^{(n)}_{10}&=\sum_{m,l}\delta_{m+l,n}\bigl(S_{1,10}^{(m)}V_{d,00}^{(l)}-V_{d,11}^{(m)}S_{1,10}^{(l)}\bigr)=-\sum_{m}V_{d,11}^{(m)}S_{1,10}^{(n-m)}\,,\\    [S_1(\tau),V_d(\tau)]^{(n)}_{01}&=\sum_{m,l}\delta_{m+l,n}\bigl(S_{1,01}^{(m)}V_{d,11}^{(l)}-V_{d,00}^{(m)}S_{1,01}^{(l)}\bigr)=\sum_{m}S_{1,01}^{(m)}V_{d,11}^{(n-m)}\,.
\end{align}
Making use of the explicit form of $S_1^{(m)}$, we arrive at
\begin{align}
S_{2}^{(n)}&=-\sum_{m}\left(\frac{1}{(U-n\omega_c)(U-(n-m)\omega_c)}V_{d,11}^{(m)}V_{od,10}^{(n-m)}-\frac{1}{(U+n\omega_c)(U+m\omega_c)}V_{od,01}^{(m)}V_{d,11}^{(n-m)}\right)\,.
\end{align}
This allows us to write the third-order correction to the Hamiltonian projected onto the singly occupied sector as
\begin{align}
    \Delta H_3(\tau)
&=\frac12 P[S_2(\tau),V_{od}(\tau)]P
=\sum_n\Delta H_3^{(n)}e^{-in\omega_c\tau}\,,\\
    \Delta H_3^{(n)}&=\frac12\sum_{r,s,t}\delta_{r+s+t,n}
V_{od,01}^{(r)}V_{d,11}^{(s)}V_{od,10}^{(t)}
\left(
\frac{1}{[U+(r+s)\omega_c](U+r\omega_c)}
+
\frac{1}{[U-(s+t)\omega_c](U-t\omega_c)}
\right)\nonumber\\
&\equiv\sum_{j}\mathcal J_{jj+1}^{(n)}\left({\bs S}_{j}\cdot{\bs S}_{j+1}-\frac{1}{4}\right)
+\mathcal K^{(n)}{\bs S}_1\cdot({\bs S}_2\times{\bs S}_3),
\end{align}
where
\begin{align}
    \mathcal J_{j,j+1}^{(n)}
&=\sum_{(a,b,c)\in{\rm Perm}(1,2,3)}
\epsilon_{abc}(2\delta_{jb}-1)C_{abc}^{(n)}\,,
\qquad
\mathcal K^{(n)}
=2i\sum_{(a,b,c)\in{\rm Perm}(1,2,3)}
C_{abc}^{(n)}\,\,,\\
C_{abc}^{(n)}
&=\frac{t_1t_2t_3}{2}
\sum_{r,s,t}\delta_{r+s+t,n}
\Big(
B_{g_a}^{(r)}B_{g_b}^{(s)}B_{g_c}^{(t)}
-B_{-g_a}^{(r)}B_{-g_b}^{(s)}B_{-g_c}^{(t)}
\Big)\left(
\frac{1}{[U+(r+s)\omega_c](U+r\omega_c)}
+\frac{1}{[U-(s+t)\omega_c](U-t\omega_c)}
\right)\,.
\end{align}
Here $a,b,c$ label bonds, $\epsilon_{123}=1$, and site indices
are understood modulo three. Both permutation sums run over
${\rm Perm}(1,2,3)$. When the photon sideband energies with appreciable weight are
small compared with $U$, they may be neglected in the denominators.
All six ordered cyclic processes then acquire the same photon
factor, depending only on $g=\sum_jg_j$. The third-order Heisenberg
corrections cancel, leaving the scalar-chirality interaction.
Returning to the laboratory frame and omitting the
spin-independent constant, we obtain:
%(strictly speaking, it is not entirely obvious whether allowing site-dependent $g_j$ would qualitatively modify the final result, but assuming a uniform $g$ does make the intermediate expressions noticeably more tractable at first sight).
\begin{align}
H_{\rm s-ph}
&\approx\sum_jJ_j{\bs S}_j\cdot{\bs S}_{j+1}
-K\sin[g(a^\dagger+a)]
{\bs S}_1\cdot({\bs S}_2\times{\bs S}_3)
+\omega_ca^\dagger a\,,
\end{align}
where $g=\sum_{j=1}^3g_j$, and $K=\frac{24t_1t_2t_3}{U^2}$ is the effective third-order exchange coupling strength.

Finally, restricting the spin dynamics to the total $S=1/2$
manifold yields the effective chirality-photon Hamiltonian
\begin{align}
H_{c-ph}
&\approx\frac12{\bs \Delta}\cdot{\bs C}
-K'\sin[g(a^\dagger+a)]C_z+\omega_ca^\dagger a\,,\\
\Delta_x&=J_1-\frac{J_2+J_3}{2},
\qquad
\Delta_y=\frac{\sqrt3}{2}(J_3-J_2),
\qquad
\Delta_z=0,
\qquad
K'=\frac{\sqrt3}{4}K\,,
\end{align}
where we have omitted an additive constant. These operators act as Pauli matrices on the chirality degree of freedom within each fixed-$S_z$ doublet of the $S=1/2$ manifold and vanish in the $S=3/2$ manifold. The resulting Hamiltonian coincides with the effective model derived in the main text and explicitly shows that, within the low-frequency approximation, the cavity affects only the three-spin ring exchange term, while the nearest-neighbor Heisenberg interaction retains its conventional form.

The derivation presented above also provides a systematic route beyond the low-frequency approximation employed in the main text. Retaining the full photon-frequency dependence of the virtual intermediate states leads to photon-dependent exchange and chirality couplings, whose quantitative consequences lie beyond the scope of the present work but provide a natural starting point for future studies of cavity-engineered orbital magnetism.

\section{Spin-orbit interaction effects}

We consider a general bond-dependent
Dzyaloshinskii–Moriya interaction
\cite{Dzyaloshinsky1958,Moriya1960}
on a triangular spin cluster,
\begin{equation}
H_{\rm DM}=\sum_{j}
\bs D_{jj+1}\cdot
\left(
\bs S_j\times \bs S_{j+1}
\right),
\end{equation}
where the vectors $\bs D_{jj+1}$ are allowed to differ from bond to bond. We assume that the isotropic exchange interaction dominates, such that the low-energy sector consists of the four-dimensional manifold with total spin $S=1/2$, while the $S=3/2$ quartet is separated by a finite (exchange) gap.

To unravel the structure of the DM interaction, we decompose the DM vectors into a uniform and a nonuniform contribution,
\begin{equation}
\bs D_{jj+1}=
{\bs D}_0
+
\delta\bs D_{jj+1},
\qquad
{\bs D}_0=
\frac{1}{3}
\left(
\bs D_{12}
+
\bs D_{23}
+
\bs D_{31}
\right),
\end{equation}
with
\begin{equation}
\delta\bs D_{12}
+
\delta\bs D_{23}
+
\delta\bs D_{31}
=0.
\end{equation}
The projection of the uniform component onto the $S=1/2$ manifold is readily obtained from the identity
\begin{equation}
P_{1/2}
\left(
\bs S_j\times \bs S_{j+1}
\right)
P_{1/2}=\frac{1}{\sqrt{3}}
C_z\,
\bs S\,,
\end{equation}
where $P_{1/2}$ denotes the projector onto the total-spin
$S=1/2$ manifold, and ${\bs S}=\sum_{j=1}^3{\bs S}_j$ is the total spin operator. Consequently,
\begin{equation}
P_{1/2}
H_{\rm DM}P_{1/2}=\sqrt{3}
C_z\,
{\bs D}_0\cdot \bs S\,.
\end{equation}
Choosing the spin quantization axis along
$\boldsymbol D_0$ gives
$H_{\rm SO}=\Delta_{\rm SO}C_zS_z$,
with $\Delta_{\rm SO}=\sqrt3|\boldsymbol D_0|$. Thus, within the low-energy manifold, the uniform DM interaction reduces to a spin-chirality coupling. The nonuniform component has a vanishing projection within the $S=1/2$ manifold because $\sum_j\delta{\bs D}_{jj+1}=0$. It nevertheless couples this manifold to the excited $S=3/2$ quartet, generating corrections of order
$\delta D^2/\Delta_{\rm ex}$, where $\Delta_{\rm ex}$ is the
exchange gap and $\delta D$ denotes the characteristic magnitude of the nonuniform DM vectors. We neglect these corrections, assuming $\delta D\ll\Delta_{\rm ex}$.

\section{Collective fluctuations and chirality polaritons}

\subsection{Effective chirality polaritons}

We consider identical molecules at zero temperature, neglecting spin–orbit coupling. The cavity coupling per molecule is
$g/\sqrt N$, so that
$\hat\Phi=g(a+a^\dagger)/\sqrt N$.
Writing $a=\sqrt N\,\alpha+d$, with the real displacement
$\alpha$ determined by mean-field minimization, gives
\begin{equation}
\hat\Phi=\Phi_0+\frac{g}{\sqrt N}(d+d^\dagger),
\qquad
\Phi_0=2g\alpha.
\end{equation}
Within a fixed spin sector, the single-molecule mean-field
Hamiltonian is
\begin{equation}
h_{\rm MF}=\frac{\Delta}{2}C_x-K'\sin\Phi_0\,C_z\,,
\end{equation}
where we rotated the in-plane chirality axes to align the splitting with \(C_x\), leaving \(C_z\) unchanged.
The ground and excited eigenstates of $h_{\rm MF}$
at the equilibrium flux $\Phi_0$ are denoted by
$|0\rangle$ and $|1\rangle$, with energies $-E$
and $E$, respectively, where
\begin{equation}
E=\sqrt{\frac{\Delta^2}{4}+K'^2\sin^2\Phi_0}.
\end{equation}
Although each molecule has a finite-dimensional chirality
Hilbert space, dilute collective excitations admit a bosonic
description. Defining the collective transition operators
\begin{equation}
\Sigma_{ab}=\sum_{j=1}^N|a\rangle_j\langle b|_j,
\end{equation}
with $a,b\in\{0,1\}$, we obtain
\begin{equation}
\left[\frac{\Sigma_{01}}{\sqrt N},
      \frac{\Sigma_{10}}{\sqrt N}\right]
=\frac{\Sigma_{00}-\Sigma_{11}}{N}
=1-\frac{2\Sigma_{11}}{N}.
\end{equation}
For an excitation population much smaller than $N$, this
commutator approaches the canonical bosonic value. Within the permutation-symmetric sector, the saturation
constraint is incorporated exactly by the
Holstein-Primakoff representation \cite{HolsteinPrimakoff40}
\begin{align}
\Sigma_{00}&=N-b^\dagger b,
&
\Sigma_{11}&=b^\dagger b,
\\
\Sigma_{10}&=b^\dagger\sqrt{N-b^\dagger b},
&
\Sigma_{01}&=\sqrt{N-b^\dagger b}\,b,
\end{align}
where $[b,b^\dagger]=1$ and the physical occupation satisfies
$b^\dagger b\leq N$. Expanding for $b^\dagger b\ll N$ gives
$\Sigma_{10}\simeq\sqrt N\,b^\dagger$ and
$\Sigma_{01}\simeq\sqrt N\,b$.
The boson describes a collective chirality excitation, while
the square roots encode the finite excitation capacity of
the molecular ensemble. In this representation,
\begin{equation}
\sum_j h_{{\rm MF},j}
=-NE+\Delta_\alpha b^\dagger b,
\qquad
\Delta_\alpha=2E.
\end{equation}
We stress that the product state $\bigotimes_j|0\rangle_j$ is the
vacuum of $b$, while $b^\dagger$ creates a symmetric
collective transition from $|0\rangle$ to $|1\rangle$. Choosing the phase of $|1\rangle$ so that
$\langle0|C_z|1\rangle$ is positive, we find
\begin{equation}
\sum_j C_{z,j}
=\frac{K'\sin\Phi_0}{E}(N-2b^\dagger b)
+\sqrt N\,\frac{\Delta}{2E}(b+b^\dagger)
+\mathcal O(N^{-1/2}),
\end{equation}
where the remainder is understood for occupations of
order unity. The mean-field energy per molecule is
$\mathcal E_{\rm MF}=\omega_c\alpha^2-E$, with the stationarity
condition
\begin{equation}
\omega_c\alpha
=\frac{gK'^2\sin\Phi_0\cos\Phi_0}{E}.
\end{equation}
Expanding the sinusoidal coupling through second order
in the fluctuations, we find that the stationarity
condition cancels the terms linear in $d+d^\dagger$. Up to an additive constant, the quadratic Hamiltonian is
\begin{equation}
H_{\rm fl}
=\omega_c d^\dagger d+A_\alpha(d+d^\dagger)^2
+\Delta_\alpha b^\dagger b
+G_\alpha(d+d^\dagger)(b+b^\dagger),
\label{eq:chirality_fluctuations}
\end{equation}
where
\begin{equation}
A_\alpha=\frac{g^2K'^2\sin^2\Phi_0}{\Delta_\alpha},
\qquad
G_\alpha=-\frac{gK'\Delta\cos\Phi_0}{\Delta_\alpha}.
\end{equation}
The quadratic cavity term originates from the curvature
of the sinusoidal Peierls coupling and must be retained
in the ordered phase. The sign of $G_\alpha$ can be changed
by redefining $b\to-b$ and does not affect the spectrum.
Defining
$\widetilde\omega_c^{\,2}=\omega_c(\omega_c+4A_\alpha)$,
diagonalization gives
\begin{equation}
\omega_\pm^2
=\frac{\widetilde\omega_c^{\,2}+\Delta_\alpha^2}{2}
\pm\frac12
\sqrt{
(\widetilde\omega_c^{\,2}-\Delta_\alpha^2)^2
+16\omega_c\Delta_\alpha G_\alpha^2
}.
\label{eq:chirality_polariton_frequencies}
\end{equation}

\subsection{Full Hubbard polaritons}

To benchmark this effective description, we construct the
Gaussian fluctuation spectrum using the complete local
Hubbard spectrum of a triangle. Let $h_{\rm Hub}(\Phi)$
denote the single-triangle electronic Hamiltonian, with
the relative distribution of the Peierls phases among
the bonds held fixed as $\Phi$ varies. At its Hubbard
mean-field saddle point $\Phi_0=2g\alpha$, we diagonalize
\begin{equation}
h_{\rm Hub}(\Phi_0)|m\rangle=E_m|m\rangle,
\qquad
\Omega_m=E_m-E_0.
\end{equation}
The states $|m\rangle$ span the complete local Hubbard
Hilbert space in the chosen symmetry sector, labeled
by $(N_{\rm el},S_{\rm tot},M_{\rm tot})$.
 Defining
\begin{equation}
V=\left.\partial_\Phi h_{\rm Hub}\right|_{\Phi_0},
\qquad
W=\left.\partial_\Phi^2h_{\rm Hub}\right|_{\Phi_0},
\end{equation}
the local Hamiltonian expands as
\begin{equation}
h_{\rm Hub}(\Phi_0+\delta\Phi)
=h_{\rm Hub}(\Phi_0)+\delta\Phi V
+\frac12(\delta\Phi)^2W+\cdots\,,
\end{equation}
where
$\delta\Phi=\frac{g}{\sqrt N}(d+d^\dagger)$. The Hubbard mean-field energy per molecule is
$\mathcal E_{\rm MF}^{\rm Hub}=\omega_c\alpha^2+E_0(\Phi_0)$.
Its stationarity condition,
\begin{equation}
\omega_c\alpha+g\langle0|V|0\rangle=0,
\end{equation}
again cancels the terms linear in the cavity fluctuation. The collective-transition construction extends to the
multilevel Hubbard spectrum \cite{KuruczMolmer10}. At leading order,
$\Sigma_{m0}\simeq\sqrt N\,b_m^\dagger$, where $b_m^\dagger$
creates the symmetric collective excitation associated
with the local transition $|0\rangle\to|m\rangle$. Here $|0\rangle$ denotes the ground state of the full
local Hubbard Hamiltonian at its own mean-field saddle,
within the chosen particle-number and spin sector.
All expectation values below are evaluated in this state,
and $V(t)$ evolves under $h_{\rm Hub}(\Phi_0)$.
The resulting quadratic Hamiltonian is
\begin{align}
H_{\rm fl}^{\rm Hub}
={}&\omega_c d^\dagger d
+A_{\rm Hub}(d+d^\dagger)^2
+\sum_{m\ne0}\Omega_m b_m^\dagger b_m+(d+d^\dagger)\sum_{m\ne0}
\left(G_m b_m^\dagger+G_m^*b_m\right),
\end{align}
with
\begin{equation}
A_{\rm Hub}=\frac{g^2}{2}\langle0|W|0\rangle,
\qquad
G_m=g\langle m|V|0\rangle.
\end{equation}
All local excited states within the chosen sector
are retained, and  transitions with vanishing matrix
elements have $G_m=0$. Equivalently, the electronic contribution can be expressed
through the zero-temperature retarded susceptibility
\begin{align}
\chi_{VV}^{R}(\omega)
&=-i\int_0^\infty dt\,
e^{i(\omega+i0^+)t}
\langle[V(t),V(0)]\rangle=\sum_{m\ne0}
\frac{2\Omega_m|\langle0|V|m\rangle|^2}
{(\omega+i0^+)^2-\Omega_m^2}.
\label{eq:Hubbard_flux_susceptibility}
\end{align}
Eliminating the collective electronic modes yields the
equation for the cavity-active excitation frequencies,
\begin{equation}
\omega^2-\omega_c^2
-2\omega_c g^2
\left[
\langle0|W|0\rangle+\chi_{VV}^{R}(\omega)
\right]=0.
\label{eq:Hubbard_polariton_kernel}
\end{equation}
Both the static curvature term and the dynamical
susceptibility are required. Their zero-frequency
combination obeys
\begin{equation}
\frac{d^2E_0}{d\Phi^2}\bigg|_{\Phi_0}
=\langle0|W|0\rangle+\chi_{VV}^{R}(0),
\end{equation}
providing a consistency check between the fluctuation
kernel and the mean-field energy curvature.

For the effective chirality Hamiltonian, these expressions
reduce to
\begin{equation}
V_\alpha=-K'\cos\Phi_0\,C_z,
\qquad
W_\alpha=K'\sin\Phi_0\,C_z,
\end{equation}
with a single active transition of energy $\Delta_\alpha$.
Consequently,
\begin{equation}
\frac{g^2}{2}\langle W_\alpha\rangle=A_\alpha,
\qquad
g^2\chi_{V_\alpha V_\alpha}^{R}(\omega)
=\frac{2\Delta_\alpha G_\alpha^2}
{(\omega+i0^+)^2-\Delta_\alpha^2}.
\end{equation}
Equation~\eqref{eq:Hubbard_polariton_kernel} then becomes
\begin{equation}
(\omega^2-\widetilde\omega_c^{\,2})
(\omega^2-\Delta_\alpha^2)
-4\omega_c\Delta_\alpha G_\alpha^2=0,
\end{equation}
recovering Eq.~\eqref{eq:chirality_polariton_frequencies}.
The Hubbard calculation therefore supplies a microscopic
benchmark that retains all local electronic transitions,
while treating the collective fluctuations at Gaussian order.

\section{Mean-field comparison and current-inferred chirality}

For the mean-field comparison in Fig.~\ref{fig2}($a$), we retain all 20 states of the three-electron Hubbard Hilbert space. We denote by $h_{\rm Hub}(\Phi)$ the electronic Hamiltonian at a dimensionless total Peierls phase $\Phi$ around the triangle. The individual bond phases are written as $\phi_j=\eta_j\Phi$, with $\sum_j\eta_j=1$ and the coefficients $\eta_j$ held fixed as $\Phi$ varies. The Hubbard mean-field displacement $\alpha_\ast$ minimizes $\mathcal E_{\rm MF}^{\rm Hub}(\alpha)=\omega_c\alpha^2+E_0(2g\alpha)$, where $E_0(\Phi)$ is the lowest eigenvalue of $h_{\rm Hub}(\Phi)$. The corresponding electronic ground state is $|0\rangle$, with equilibrium phase $\Phi_\ast=2g\alpha_\ast$.

To characterize the orbital response, we introduce the operator conjugate to the total Peierls phase,
\[
\hat{\mathcal I}_{\rm Hub}(\Phi)
\equiv
\frac{\partial h_{\rm Hub}(\Phi)}{\partial\Phi}
=
i\sum_{j,\sigma}\eta_jt_j
\left[
e^{i\eta_j\Phi}c^\dagger_{j\sigma}c_{j+1\sigma}
-
e^{-i\eta_j\Phi}c^\dagger_{j+1\sigma}c_{j\sigma}
\right].
\]
This operator has units of energy. With the bond-current convention used in the main text, the physical current on bond $j$ is $\hat I_j=(e/\hbar)\partial_{\phi_j}h_{\rm Hub}$, so that $\hat{\mathcal I}_{\rm Hub}=(\hbar/e)\sum_j\eta_j\hat I_j$. For equally distributed phases, $\eta_j=1/3$, this is proportional to the average of the three bond-current operators. In a stationary electronic state, charge conservation requires equal current expectation values on all three oriented bonds. Consequently, the physical circulating current is $(e/\hbar)\mathcal I_{\rm Hub}$, where
\[
\mathcal I_{\rm Hub}(\Phi_\ast)
=
\langle0|
\hat{\mathcal I}_{\rm Hub}(\Phi_\ast)
|0\rangle .
\]

The third-order effective Hamiltonian contains the flux-dependent term $-K'\sin\Phi\,C_z$, giving the low-energy current relation $\mathcal I_{\rm eff}=-K'\cos\Phi\,\langle C_z\rangle$. We therefore define the current-inferred Hubbard chirality as
\[
\langle C_z^{(I)}\rangle
=
-\frac{\mathcal I_{\rm Hub}(\Phi_\ast)}
{K'\cos\Phi_\ast},
\qquad
K'=\frac{6\sqrt{3}\,t_1t_2t_3}{U^2},
\]
where $\cos\Phi_\ast\neq0$. The superscript $(I)$ emphasizes that this quantity is inferred from the circulating current through the leading ring-exchange relation. Higher-order strong-coupling corrections are retained in the full Hubbard current and can produce deviations from the effective-model chirality.

We compare $\langle C_z^{(I)}\rangle$ with $\langle C_z\rangle$ evaluated at the effective chirality model's independently determined mean-field minimum. Both observables within each model therefore refer to the same self-consistent state. Finally, the Hellmann-Feynman relation gives $\partial_\Phi E_0=\mathcal I_{\rm Hub}$, so the Hubbard mean-field stationarity condition reads
\[
\omega_c\alpha_\ast
+
g\,\mathcal I_{\rm Hub}(\Phi_\ast)
=0.
\]
This relation provides a numerical consistency check between the photon displacement and the circulating current.

\section{Numerical convergence}

For Fig.~\ref{fig3}($a$), we locate the finite-size
pseudocritical coupling from the maximum slope of
the parity-even photon amplitude $\alpha_{\rm fl}$
defined in the main text. This diagnostic is motivated
by photonic observables used in finite-size studies
of the Dicke transition \cite{EmaryPRE03,Vidal_2006}.
The subtraction removes the vacuum contribution
$\langle X^2\rangle_{\rm vac}=1$, and the normalization
gives $\alpha_{\rm fl}=|\alpha|$ for a real coherent
state $|\sqrt N\alpha\rangle$. At $N=200$, we checked photon-cutoff and coupling-grid
convergence at $\Delta=0.0205$, $0.15$, and $0.30$,
along the microscopic parameter trajectory used in
Fig.~\ref{fig3}($a$).
Comparing photon cutoffs
$n_{\rm ph}=50$, $70$, and $100$ changes the
maximum-slope pseudocritical couplings by less than
$3\times10^{-7}$.
Refining the coupling grid from $80$ to $160$ intervals
shifts these couplings by less than $0.04\%$.
The scans cover $g\in[0.70g_c^{\rm MF},1.45g_c^{\rm MF}]$.
These checks establish numerical stability at the
tested finite-size points.

\begin{table}[h]
\centering
\begin{tabular}{cc|cc}
\hline\hline
$n_{\rm ph}$ & $g_c(N)$ &
Grid intervals & $g_c(N)$ \\
\hline
50  & 0.86500233 & 40  & 0.86725648 \\
70  & 0.86500241 & 80  & 0.86527168 \\
100 & 0.86500231 & 160 & 0.86500231 \\
\hline\hline
\end{tabular}
\caption{Photon-cutoff and coupling-grid checks at
$N=200$ and $\Delta=0.0205$.
The cutoff comparison uses the 160-interval grid;
the grid comparison uses $n_{\rm ph}=100$.}
\end{table}
All 18 sampled values of $\Delta$ were evaluated
through $N=240$. The extrapolated estimates in
Fig.~\ref{fig3}($a$) are obtained from unweighted fits
to
\begin{equation}
g_c(N)=g_c(\infty)+rN^{-2/3},
\end{equation}
using $N=\{120,160,200,240\}$, except at
$\Delta=0.15$ and $0.30$, where we use
$N=\{200,240,320,400\}$.
The latter gives $g_c(\infty)=1.78003$ and $2.11032$,
respectively, compared with mean-field values
$1.78174$ and $2.10819$.
Increasing the photon cutoff from $70$ to $100$
changes $g_c(400)$ by less than $4\times10^{-8}$
at these points. We assess extrapolation sensitivity by varying
the fitting window and including a subleading
term $sN^{-4/3}$ in fits with at least four sizes.
All plotted estimates use the leading four-size
fits, with unconstrained intercepts.
The variation between fits characterizes
extrapolation sensitivity rather than a
statistical confidence interval.

For the biased entropy curves in Fig.~\ref{fig3}($b$),
we include $H_{\rm bias}=h\sqrt N(a+a^\dagger)$
with $h=10^{-4}$ and compare photon cutoffs $n_{\rm ph}=60$, $80$, and $100$ for $N=40$, $100$, and $200$.
The maximum absolute entropy differences over the
sampled curves are below $1.5\times10^{-8}$,
and the sampled peak locations remain unchanged.
Their resolution is limited by the coupling-grid
spacing, $\delta g\simeq0.0156$.

For the Hubbard-chirality comparison in
Fig.~\ref{fig3}($c$), we use $t_1=t_2=1$, $t_3=1.05$,
$U=30$, and $\omega_c=0.1$, with a selecting field
$H_{\rm bias}=h\sqrt N X$, where $h=10^{-4}$.
Both models are evaluated at $49$ equally spaced
coupling points over $0\leq g\leq2$.
We plot $|\langle X\rangle|/(2\sqrt N)$.
The plotted curves use photon cutoffs
$n_{\rm ph}=40,20,20,12$ for $N=1,2,3,4$,
respectively. Increasing the cutoffs from $20$ to $40$ for $N=1$
and from $10$ to $20$ for $N=2,3$ changes the
displacement by less than $4\times10^{-8}$.
For $N=4$, increasing the cutoff from $8$ to $12$
changes it by less than $3\times10^{-8}$.
The differences between the Hubbard and chirality
curves therefore persist after photon-cutoff refinement.

For Fig.~\ref{fig3}($d$), we consider $N=20$ in a
fixed, fully polarized spin sector, with
a selecting-field magnitude $h=10^{-5}$.
Increasing the photon cutoff from $n_{\rm ph}=30$
to $70$ changes the plotted displacement by at
most $3.4\times10^{-10}$.  Refining the coupling-grid spacing from $0.01$ to
$0.005$ shifts the maximum-slope location by at most
$6.7\times10^{-5}$.
At zero SOI, this location remains sensitive to the
selecting-field strength: changing $h$ by a factor
of ten shifts it by approximately $0.12\text{--}0.13$.
The panel therefore characterizes the finite-size
response at $h=10^{-5}$ rather than a thermodynamic
phase boundary.

\section{Temperature effects on the superradiant phase transition}

We now include the projected DM interaction and extend the
mean-field treatment to finite temperature. We choose the spin quantization axis along $\mathbf D_0$
and retain $\Delta_{\rm SO}=\sqrt{3}|\mathbf D_0|$
as defined above. With the collective scaling
introduced above, $\langle a\rangle=\sqrt N\,\alpha$, the
single-molecule Hamiltonian reads
\begin{equation}
h_{\rm MF}(\alpha)
=\frac{\Delta}{2}C_x
-K'\sin(2g\alpha)C_z
+\Delta_{\rm SO}C_zS_z.
\label{eq:Heff_appendix}
\end{equation}
Here $\Delta_{\rm SO}$ is the coefficient of the projected
spin-chirality coupling, rather than the full molecular
excitation gap. Since $S_z$ is conserved, the Hamiltonian separates into two chirality blocks labeled by $\sigma=2S_z=\pm1$:
\begin{equation}
h_\sigma(\alpha)
=
\begin{pmatrix}
q_\sigma(\alpha) & \Delta/2\\
\Delta/2 & -q_\sigma(\alpha)
\end{pmatrix},
\qquad
q_\sigma(\alpha)
=-K'\sin(2g\alpha)+\sigma\frac{\Delta_{\rm SO}}{2}.
\end{equation}
The corresponding eigenvalues are $\pm E_\sigma(\alpha)$, with
\begin{equation}
E_\sigma(\alpha)
=
\sqrt{
\left(\frac{\Delta}{2}\right)^2
+q_\sigma^2(\alpha)
}.
\end{equation}
The chirality excitation gap within each spin sector is
$2E_\sigma(\alpha)$.

We assume $k_BT\ll\Delta_{\rm ex}$, so that thermal
occupation of the excited $S=3/2$ quartet can be neglected. Then, tracing over all four states of the projected molecular
Hilbert space gives the equilibrium free energy per molecule,
\begin{equation}
\mathcal F(\alpha)
=\omega_c\alpha^2
-\frac{1}{\beta}\ln\mathcal Z_{\rm el}(\alpha),
\qquad
\mathcal Z_{\rm el}(\alpha)
=2\sum_{\sigma=\pm1}\cosh[\beta E_\sigma(\alpha)],
\label{eq:free_energy_full}
\end{equation}
where $\beta=(k_BT)^{-1}$. This trace assumes equilibrium
populations of both spin sectors. Because
$E_\sigma(-\alpha)=E_{-\sigma}(\alpha)$, the free energy is even
in $\alpha$ and is analytic at finite temperature.

\subsection{Unconstrained spin polarization}

To determine the local stability of the normal state, we
introduce
\begin{equation}
E_{\rm ref}=E_\sigma(0)
=\sqrt{
\left(\frac{\Delta}{2}\right)^2
+\left(\frac{\Delta_{\rm SO}}{2}\right)^2
}.
\end{equation}
Expanding around $\alpha=0$ yields
\begin{equation}
\mathcal F(\alpha)
=\mathcal F(0)
+\left[
\omega_c-2K'^2g^2\chi_{\rm eff}(T,\Delta_{\rm SO})
\right]\alpha^2
+\mathcal O(\alpha^4),
\end{equation}
where
\begin{equation}
\chi_{\rm eff}(T,\Delta_{\rm SO})
=
\frac{(\Delta/2)^2}{E_{\rm ref}^3}\tanh(\beta E_{\rm ref})
+\beta\frac{(\Delta_{\rm SO}/2)^2}{E_{\rm ref}^2}.
\label{eq:chi_eff_SO}
\end{equation}
Here $\chi_{\rm eff}$ is the static susceptibility to a field
coupled to $C_z$. Its first term describes virtual transitions
between the chirality levels, while the second arises from
redistribution between the two spin sectors and has a Curie-like temperature dependence. The normal state loses local stability at
\begin{equation}
g_c(T,\Delta_{\rm SO})
=
\sqrt{
\frac{\omega_c}
{2K'^2\chi_{\rm eff}(T,\Delta_{\rm SO})}
}.
\label{eq:gc_SO_finite_T}
\end{equation}
This is the equilibrium transition coupling when the transition
is continuous. For $\Delta_{\rm SO}=0$ and $T\rightarrow0$,
it reduces to
\begin{equation}
g_c(0,0)=\frac{\sqrt{\omega_c\Delta}}{2|K'|}.
\end{equation}
For nonzero $\Delta_{\rm SO}$, the Curie contribution instead
dominates as $T\rightarrow0$, giving
\begin{equation}
g_c(T,\Delta_{\rm SO})
\simeq
\frac{E_{\rm ref}}{|K'|\Delta_{\rm SO}}
\sqrt{2\omega_c k_BT}.
\end{equation}
At strictly zero temperature, minimizing over the two spin
sectors gives
\begin{equation}
\mathcal F_{T=0}(\alpha)
=
\omega_c\alpha^2
-\sqrt{
\left(\frac{\Delta}{2}\right)^2
+\left[
\frac{\Delta_{\rm SO}}{2}
+\left|K'\sin(2g\alpha)\right|
\right]^2
}.
\label{eq:F_zero_T_SO}
\end{equation}
For $\Delta_{\rm SO}>0$ and $g\geq0$, its small-displacement
expansion is
\begin{equation}
\mathcal F_{T=0}(\alpha)-\mathcal F_{T=0}(0)
=
-\frac{|K'|g\Delta_{\rm SO}}{E_{\rm ref}}|\alpha|
+\mathcal O(\alpha^2).
\label{eq:F_cusp_SO}
\end{equation}
The negative linear cusp dominates the quadratic cavity cost,
so any nonzero coupling destabilizes $\alpha=0$ and favors
symmetry-related displaced minima. Finite temperature rounds the cusp and restores a finite normal-state instability
threshold. The sinusoidal and linearized couplings give the same local
instability condition because
$\partial_\alpha\sin(2g\alpha)|_{\alpha=0}=2g$.
Away from the normal state, their free energies differ, and the
full sinusoidal dependence determines the equilibrium
displacement and the fluctuation spectrum.

\subsection{Fixed magnetization at zero temperature}

If the total spin polarization is constrained to vanish,
the two spin sectors have equal populations. Their
ground-state energies are averaged before minimizing
over the cavity displacement:
\begin{equation}
\mathcal F_{\rm fixed}(\alpha)
=\omega_c\alpha^2-\frac12[E_+(\alpha)+E_-(\alpha)].
\end{equation}
The resulting energy remains even in $\alpha$ and,
for $\Delta>0$, is analytic near $\alpha=0$. Expanding the ground-state energy to quadratic order gives
\begin{equation}
\mathcal{F}_{\rm fixed}(\alpha)
=
\mathcal{F}_{\rm fixed}(0)
+
\left[
\omega_c
-
\frac{4g^2K'^2\Delta^2}
{(\Delta^2+\Delta_{\rm SO}^2)^{3/2}}
\right]
\alpha^2
+
\mathcal O(\alpha^4).
\end{equation}
The instability therefore occurs at
\begin{equation}
g_c(\Delta_{\rm SO})
=
g_c(0)
\left[
1+
\left(
\frac{\Delta_{\rm SO}}{\Delta}
\right)^2
\right]^{3/4}\,.
\end{equation}
This is the normal-state instability threshold and equals 
the transition coupling when the transition is continuous.

\section{Experimental estimates for the orbital light-matter coupling}

The zero-point current fluctuations of a superconducting LC resonator are given by
\cite{Girvin2014}
\begin{equation}
I_{\rm zpf}=\sqrt{\frac{\hbar\omega_c}{2L}}.
\end{equation}
Approximating the inductive element as a circular loop of radius $R$, the
magnetic field at its center follows from the Biot-Savart law as
\begin{equation}
B_{\rm zpf}
\simeq
\frac{\mu_0 I_{\rm zpf}}{2R}.
\end{equation}
A molecular triangle of area $A_\triangle$, oriented approximately parallel
to the loop, then experiences the zero-point flux
\begin{equation}
\Phi_{\rm zpf}
\simeq
B_{\rm zpf}A_\triangle,
\end{equation}
leading to the dimensionless Peierls coupling
\begin{equation}
g_i=2\pi\frac{\Phi_{{\rm zpf},i}}{\Phi_e},
\qquad
\Phi_e=\frac{h}{e}.
\end{equation}
For an ensemble of molecular triangles, the collective coupling to the
resonator mode is characterized by
\begin{equation}
g=\sqrt{\sum_{i=1}^{N}g_i^2}.
\end{equation}
Here $\Phi_{{\rm zpf},i}$ is the physical zero-point
magnetic flux through triangle $i$, while $g_i$
is the corresponding dimensionless Peierls-phase
amplitude. The phase operator entering its
ring-exchange interaction is
\begin{equation}
\hat\Phi_{{\rm cav},i}=g_i(a+a^\dagger).
\end{equation}
For approximately uniform coupling,
$g_i\simeq g_{\rm mol}$ and $g\simeq g_{\rm mol}\sqrt N$.
Consequently,
$\hat\Phi_{{\rm cav},i}
=\hat\Phi_{\rm cav}/\sqrt N$,
with $\hat\Phi_{\rm cav}=g(a+a^\dagger)$,
recovering the collective convention used in the main text. 
%For approximately uniform coupling, $g_i\simeq g_{\rm mol}$ and $g\simeq g_{\rm mol}\sqrt N$.

%in this case, $G_N$ corresponds to the collectively scaled parameter $g$ used above.

To obtain a conservative estimate, we consider a circular superconducting loop
with representative parameters
\begin{equation}
R\sim25~\mu{\rm m},
\qquad
L\sim1~{\rm nH},
\qquad
\omega_c\sim2\pi\times5~{\rm GHz},
\end{equation}
together with a molecular area $A_\triangle\sim1~{\rm nm}^2$. These parameters give:
\begin{equation}
I_{\rm zpf}=\sqrt{\frac{\hbar\omega_c}{2L}}
\sim4\times10^{-8}~{\rm A},
\end{equation}
and hence
\begin{equation}
B_{\rm zpf}
\sim10^{-9}~{\rm T}.
\end{equation}
The corresponding single-molecule Peierls coupling is therefore of order
\begin{equation}
g_{\rm mol}\sim10^{-12}\,.
\end{equation}
For an ensemble containing $N\sim10^{12}$ molecular triangles, this yields
\begin{equation}
g\simeq g_{\rm mol}\sqrt N\sim10^{-6}\,.
\end{equation}
Assuming a molecular number density of order
$1~{\rm nm}^{-3}$, $N\sim10^{12}$ corresponds to a volume
of approximately $10^3~\mu{\rm m}^3$. This could be accommodated in platelets with lateral
dimensions of $10\times10$ and $20\times20~\mu{\rm m}^2$
and thicknesses of approximately $10$ and
$2.5~\mu{\rm m}$, respectively. The homogeneous-field approximation should
be assessed over this sample volume. In the normal phase, the collective transition coupling is
\begin{equation}
\Delta_{\rm cav}
=|K'|g\,|\langle1|C_z|0\rangle|.
\end{equation}
For the SOI-free model, $|\langle1|C_z|0\rangle|=1$,
giving $\Delta_{\rm cav}=|K'|g$.  For the illustrative range
\begin{equation}
\frac{K'}{h}\sim10\text{--}100~{\rm GHz}\,,
\end{equation}
the circular-loop estimate gives
\begin{equation}
\frac{\Delta_{\rm cav}}{h}
\sim
10\text{--}100~{\rm kHz}.
\end{equation}
This corresponds to an equivalent temperature scale
\begin{equation}
\frac{\Delta_{\rm cav}}{k_B}
\sim
0.5\text{--}5~\mu{\rm K}.
\end{equation}
For comparison, molecular electron-spin ensembles can
exhibit MHz-scale inhomogeneous broadening
\cite{LenzAM21}, although the linewidth of the chirality
transition considered here requires a separate
material-specific assessment. These estimates provide a baseline for a simple circular-loop
geometry. Increasing the collective orbital coupling requires
optimizing the zero-point magnetic field over the molecular
sample. Narrow inductive constrictions and other field-focusing structures offer possible routes \cite{GimenoACSNano20}, but their benefits must be evaluated through $g=\sqrt{\sum_i g_i^2}$, taking into account the spatial field profile and the resulting inductance of the resonator. If the collective coupling itself could be enhanced by a factor of $10$--$100$, the estimated interaction scale would increase to approximately $0.1$--$10~{\rm MHz}$. Whether this permits resolved chirality polaritons depends on the cavity and chirality-transition linewidths.

\end{document}